\documentclass[12pt]{iopart}
\usepackage{iopams}
\usepackage{graphicx}
\usepackage{dcolumn}
\usepackage{longtable}
\usepackage{amssymb}

\begin{document}

\title
[Bohr Hamiltonian with a deformation-dependent mass term]
{Bohr Hamiltonian with a deformation-dependent mass term: physical meaning of the free parameter}

\author
{Dennis Bonatsos$^1$, N Minkov$^2$ and D Petrellis$^3$}

\address
{$^1$ Institute of Nuclear and Particle Physics, National Center for Scientific Research 
``Demokritos'', GR-15310 Aghia Paraskevi, Attiki, Greece}

\address
{$^2$ Institute of Nuclear Research and Nuclear Energy, Bulgarian Academy of Sciences, 72 Tzarigrad Road, 1784 Sofia, Bulgaria}

\address
{$^3$ Department of Physics, University of Istanbul, 34134 Vezneciler, Istanbul, Turkey}

\eads{\mailto{bonat@inp.demokritos.gr}, \mailto{nminkov@inrne.bas.bg}, \mailto{dimitrios.petrellis@istanbul.edu.tr}}

\begin{abstract}
Embedding of the 5-dimensional (5D) space of the Bohr Hamiltonian with a deformation-dependent mass (DDM) into a 6-dimensional (6D) space shows that the free parameter in the dependence of the mass on the deformation is connected to the curvature 
of the 5D space, with the special case of constant mass corresponding to a flat 5D space. Comparison of the DDM Bohr Hamiltonian to the 5D classical limit of Hamiltonians of the 6D interacting boson model (IBM), shows that the DDM parameter is proportional to the strength of the pairing interaction in the U(5) (vibrational) symmetry limit, while it is proportional to the quadrupole-quadrupole interaction in the SU(3) (rotational) symmetry limit, 
and to the difference of the pairing interactions among $s$, $d$ bosons and $d$ bosons alone in the O(6) ($\gamma$-soft) limit.
The presence of these interactions leads to a curved 5D space in the classical limit of IBM, in contrast 
to the flat 5D space of the original Bohr Hamiltonian, which is made curved by the introduction of the deformation-dependent mass.

\end{abstract}



\maketitle
      
\section{Introduction}

The advent of supersymmetric quantum mechanics (SUSYQM) \cite{SUSYQM1,SUSYQM} has led to exact solutions of Hamiltonians with
position-dependent masses (PDM) \cite{QT,Bagchi}. Position-dependent effective masses have been widely used \cite{QT,Bagchi} in compositionally graded crystals, semiconductor theory, quantum dots, quantum liquids, liquid crystals and metal clusters. In this paper we consider the use of
a position-dependent effective mass in the structure of atomic nuclei, showing that this is not a mathematical trick introducing an extra parameter,
but a necessity arising from concrete physical requirements. 

Collective effects in atomic nuclei are described in two alternative frameworks, the collective model 
of Bohr and Mottelson \cite{Bohr,BM} and the interacting boson model (IBM) of Arima and Iachello \cite{IA}. 
A major drawback of the collective model of Bohr and Mottelson has been over the years  the rapid increase of the nuclear moments of inertia as a function of the nuclear deformation \cite{Ring}. A possible solution to this problem has been suggested \cite{DDM} recently by allowing the nuclear mass to depend on the deformation,
taking advantage of techniques developed for Hamiltonians with masses dependent on the coordinates 
\cite{QT,Bagchi}, with exact analytical solutions obtained for certain classes of potentials through the use 
of supersymmetric quantum mechanics (SUSYQM) \cite{SUSYQM1,SUSYQM}. Such a solution has been obtained \cite{DDM} 
for the deformation-dependent mass (DDM) Bohr Hamiltonian 
\cite{Bohr} with a Davidson potential \cite{Dav}, in which the mass depends on the deformation through a specific function 
containing a free parameter, $a$. It is the purpose of the present work to examine the physical relevance of this free parameter.
This will be achieved  by embedding \cite{Georg} the 5-dimensional (5D) space of the Bohr Hamiltonian into a 6-dimensional (6D) space,
thus connecting the DDM free parameter to the curvature of the 5D space, a procedure of general applicability in Hamiltonians 
using position-dependent masses. 

On the other hand, the algebraic interacting boson model (IBM) \cite{IA}, which is known to possess an overall U(6) symmetry in a 6D space, 
is also known to have a classical limit \cite{GKPRL,DSI,GKNPA,vanRoos}, corresponding to a 5D space. Comparison of the DDM Bohr Hamiltonian 
to the classical limit of the most general IBM Hamiltonian, as well as to the special cases corresponding to the three limiting 
symmetries [U(5) (vibrational), SU(3)(rotational) and O(6) ($\gamma$-unstable)] of the IBM, results in connecting the DDM free 
parameter to certain IBM parameters, thus attributing to it a specific physical meaning.   

Before proceeding, two comments are in place.

1) The need for different mass parameters in the various low-lying nuclear collective bands, 
and even for a mass tensor within the Bohr Hamlitonian, has been pointed out by Jolos and von Brentano \cite{Jolos1,Jolos2,Jolos3}. This method has been recently extended to the description of odd nuclei
\cite{Erm1,Erm2}.  

2) The Davidson potential has also been used in the framework of the algebraic collective model \cite{Rowe735,Rowe753,Welsh}, 
which allows rapidly converging numerical calculations for nuclei of any shape.   

The DDM Bohr Hamiltonian will be briefly presented in Section 2, while in Section 3 its embedding into a 6D space  
will be developed. The classical limit of the IBM Hamiltonian  will be compared to the DDM Bohr Hamiltonian 
in Section 4, while in Section 5 the results will be discussed. 

\section{The DDM Bohr Hamiltonian} 

The original Bohr Hamiltonian \cite{Bohr} is
$$
H_B = -{\hbar^2 \over 2B} \left[ {1\over \beta^4} {\partial \over \partial 
\beta} \beta^4 {\partial \over \partial \beta} + {1\over \beta^2 \sin 
3\gamma} {\partial \over \partial \gamma} \sin 3 \gamma {\partial \over 
\partial \gamma} \right.  $$
\begin{equation}\label{eq:e1}
\left. - {1\over 4 \beta^2} \sum_{k=1,2,3} {Q_k^2 \over \sin^2 
\left(\gamma - {2\over 3} \pi k\right) } \right] +V(\beta,\gamma),
\end{equation}
where $\beta$ and $\gamma$ are the usual collective coordinates
($\beta$ being a deformation coordinate measuring departure from spherical shape, 
and $\gamma$ being an angle measuring departure from axial symmetry), while
$Q_k$ ($k=1$, 2, 3) are the components of angular momentum in the intrinsic 
frame, and $B$ is the mass parameter, which is usually considered as being a constant.  

In the DDM case, the mass is assumed to be a function of the deformation 
\begin{equation}\label{mass}
B(\beta)=\frac{B_0}{(f(\beta))^2}, 
\end{equation}
where $B_0$ is a constant.

The Schr\"odinger equation for the Bohr Hamiltonian $H$ with deformation-dependent mass (DDM) has the form \cite{DDM}
$$
H \Psi = \left[ 
-{1\over 2} {\sqrt{f}\over \beta^4} {\partial \over \partial \beta} 
\beta^4 f {\partial \over \partial \beta} \sqrt{f} 
 -{f^2 \over 2 \beta^2 \sin 3\gamma} {\partial \over \partial \gamma} 
\sin 3\gamma {\partial \over \partial \gamma} \right.  
$$
\begin{equation}\label{eq:mBohr}
  \left. + {f^2\over 8 \beta^2} 
\sum_{k=1,2,3} {Q_k^2 \over \sin^2\left(\gamma -{2\over 3} \pi k \right)}
+ v_{eff} \right] \Psi = \epsilon \Psi,
\end{equation}  
where reduced energies $\epsilon = B_0 E/\hbar^2$ and reduced potentials
$v= B_0 V/\hbar^2$ have been used, the effective potential being 
\begin{equation}
v_{eff}= v(\beta,\gamma)+ {1\over 4 } (1-\delta-\lambda) f \nabla^2 f 
+ {1\over 2} \left({1\over 2} -\delta\right) \left( {1\over 2} -\lambda\right)
(\nabla f)^2 ,
\end{equation}
where $\delta$ and $\lambda$ are free parameters, further discussed in Ref. \cite{DDM}.

In order to achieve exact separation of variables \cite{Wilets,Fortunato}, potentials of the form
\begin{equation}
v(\beta,\gamma)=u(\beta)+{f^2 \over \beta^2} w(\gamma) 
\end{equation}
have been considered. In the place of $u(\beta)$, the Davidson \cite{Dav} and Kratzer \cite{Kratzer} potentials 
have been used \cite{DDM,DDMK}. For $\gamma$-unstable nuclei one has $w(\gamma)=0$, while for prolate 
deformed nuclei a harmonic oscillator potential centered around $\gamma=0$ is used, and for triaxial nuclei 
a harmonic oscillator potential centered around $\gamma=\pi/6$ is used. The potential $w(\gamma)$ is playing 
no role in what follows, since it always appears within the effective potential $v_{eff}$, which contains 
no derivatives with respect to the variables $\beta$ and $\gamma$. It should be noticed that for both  
the Davidson and the Kratzer potentials, the ``radial'' equation (the equation involving the variable $\beta$)
is acquiring a common form \cite{DDM,DDMK} for all three different cases of $w(\gamma)$ mentioned above. 

In the case of the Davidson potential \cite{Dav}
\begin{equation} \label{Davpot}
u(\beta)=\beta^2 + {\beta_0^4\over \beta^2},
\end{equation}
where $\beta_0$ is a parameter indicating the position of the minimum 
of the potential, the deformation 
function has the special form \cite{DDM} 
\begin{equation}\label{fbeta}
f(\beta)=1+a \beta^2, \qquad a \ll 1,
\end{equation}
where $a$ is a free parameter. It is the physical content 
of this parameter which will be considered in the present study. 

Performing the derivation in the first term of Eq. (\ref{eq:mBohr}),
the Hamiltonian in the case of the Davidson potential takes the form 
$$  H \Psi = \left[
 - {2\over \beta}(1+a\beta^2) (1+2 a \beta^2){\partial \over \partial \beta} 
 -{1\over 2} (1+a \beta^2)^2 {\partial ^2  \over \partial \beta^2} 
 -{(1+a\beta^2)^2 \over 2 \beta^2 \sin 3\gamma} {\partial \over \partial \gamma} 
\sin 3\gamma {\partial \over \partial \gamma} \right. $$
 \begin{equation}
\label{mBohr2}
\left.   + {(1+a\beta^2)^2\over 8 \beta^2} 
\sum_{k=1,2,3} {Q_k^2 \over \sin^2\left(\gamma -{2\over 3} \pi k \right)}
+ v'_{eff} \right] \Psi = \epsilon \Psi,
\end{equation}  
where 
\begin{equation}
v'_{eff}= v_{eff} -\left( {5\over 2} a  + 3 a^2 \beta^2 \right) .  
\end{equation}
This result will be used in Section 4 for comparison to the classical limit
of the most general IBM Hamiltonian.  

\section{Embedding of the DDM Bohr Hamiltonian} 

\subsection{Embedding from two into three dimensions}

In order to clarify the notions of curved space, embedding into higher dimensions, and conformal factor,
we start with some simple considerations.

The surface of a sphere is a 2D (two-dimensional) curved space of constant curvature, 
on which the Pythagorean theorem is not valid. 
The Pythagorean theorem is restored if we add
one more dimension, going to the familiar 3D space, 
i.e., embedding the 2D sphere in a (2+1)D space, 
in which the sphere is described by the constraint
\begin{equation}\label{x3}
x_1^2 + x_2^2 + x_3^2 = R^2, 
\end{equation}
where $R$ is the radius of the sphere,
while the length element in cartesian coordinates is 
\begin{equation}
dl^2 = dx_1^2 + dx_2^2 + dx_3^2. 
\end{equation}
Using Eq. (\ref{x3}) we can eliminate the 3rd  coordinate, obtaining 
\begin{equation}
dl^2 = dx_1^2 + dx_2^2 +{(x_1 dx_1 + x_2 dx_2)^2 \over R^2 -x_1^2 -x_2^2}. 
\end{equation}
Replacing the cartesian coordinates of the 2D euclidean space by polar ones 
\begin{equation}
x_1= r \cos\theta, \qquad x_2 = r \sin\theta,
\end{equation}
this can be rewritten as 
\begin{equation}\label{l2}
dl^2= dr^2 \left( 1-{r^2 \over R^2}\right)^{-1}  + r^2 d\theta^2.
\end{equation}
Performing the conformal transformation (i.e., a transformation 
not affecting the angles, but only modifying the radial coordinate) 
\begin{equation}\label{rconf}
r = r_1 \left({1+{r_1^2\over 4 R^2}}\right)^{-1},
\end{equation}
the length element takes the form 
\begin{equation}\label{lconf2}
dl^2 = \left( 1 + {r_1^2 \over 4 R^2}\right)^{-2} \left( dr_1^2 + r_1^2 d\theta^2  \right),
\end{equation}
which is proportional to the euclidean expression 
\begin{equation}\label{eucl2}
\bar{dl}^2 = dr_1^2 + r_1^2 d\theta^2 ,
\end{equation}
differing only by the conformal factor in front of it in Eq. (\ref{lconf2}). 

Following the same steps in the embedding from three into four dimensions \cite{Robertson,Landau} and using 
spherical coordinates one obtains 
\begin{equation}\label{lconf3}
dl^2 = \left( 1 + {r_1^2 \over 4 R^2}\right)^{-2} \left( dr_1^2 + r_1^2 (d\theta^2 +  \sin^2\theta d\phi^2) \right),
\end{equation}
which is proportional to the euclidean expression
\begin{equation}\label{eucl3}
\bar{dl}^2 = dr_1^2 + r_1^2 (d\theta^2 +  \sin^2\theta d\phi^2) .
\end{equation} 

\subsection{Stereographic projection}

Through a stereographic projection a point $(x_1,x_2,x_3)$ on the surface of a sphere is projected 
onto a point $(X_1,X_2)$ on the plane passing through the south pole of the sphere and being perpendicular
to the axis connecting the poles of the sphere. The projection is made through a line starting 
from the north pole of the sphere, passing through the $(x_1,x_2,x_3)$ point and ending at the $(X_1,X_2)$ point.
This is schematically illustrated in Fig.~1, where the plane formed by the axes 2 and 3 is depicted.   

%FIGURE 1
\begin{figure*}[hbt]
\includegraphics[width=1\textwidth]{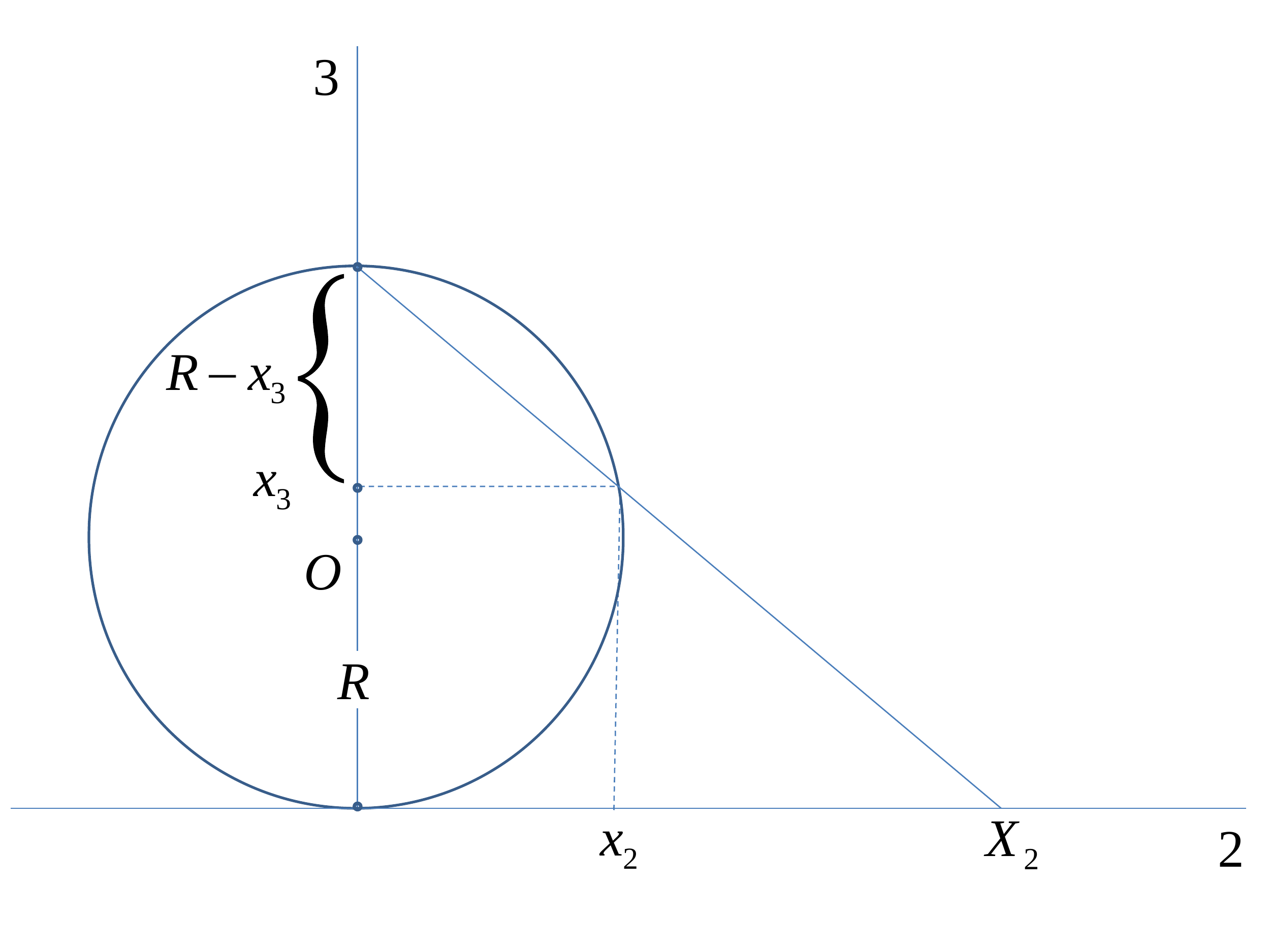}
\caption{(Color online) 
Schematic illustration (two-dimensional intersection) of the stereographic projection of a point $(x_1,x_2,x_3)$ 
on the surface of a sphere of radius $R$ onto a point $(X_1,X_2)$ on a plane passing through the south pole of the sphere and being perpendicular to the axis connecting the poles of the sphere. The intersection of the sphere with the plane formed by the axes 2 and 3 is shown. The three axes are labeled by 1, 2, 3, while distances on them are measured starting 
from the center of the sphere, $O$. The similarity between the two triangles depicted, leads directly to the second equation in (\ref{similarity}). }
\end{figure*}

If $R$ is the radius of the sphere, and the north pole is located on the positive $x_3$ axis,
from simple geometrical considerations one sees from Fig.~1 that 
\begin{equation}\label{similarity}
{x_1 \over R-x_3}= {X_1 \over 2 R}, \qquad {x_2 \over R-x_3}= {X_2 \over 2 R}. 
\end{equation}
Solving the system of these two equations plus Eq. (\ref{x3}) for $x_1$, $x_2$, $x_3$ we find
\begin{equation}
x_1= {X_1 \over D}, \quad x_2=  {X_2 \over D},\quad 
 x_3= {R\over D}  \left(-1 + { X_1^2 +  X_2^2 \over 4 R^2} \right), \quad 
D=1 + { X_1^2 + X_2^2 \over 4 R^2 }. 
\end{equation}
Using cylindrical coordinates for the point on the sphere 
\begin{equation}
x_1= r \cos\theta , \qquad x_2 = r \sin\theta, \qquad x_3=x_3,
\end{equation}
and polar coordinates on the plane passing through the south pole 
\begin{equation}
 X_1 = r_1 \cos\Theta, \qquad X_2 = r_1 \sin\Theta, 
\end{equation}
it turns out that 
\begin{equation}
r= {r_1 \over 1+ {r_1^2 \over 4 R^2}}, \qquad x_3= R {-1+{r_1^2\over 4 R^2} \over 1+{r_1^2\over 4 R^2}} ,\qquad \theta =\Theta. 
\end{equation}
The expression for $r$ coincides with the one given in Eq. (\ref{rconf}). 
We therefore see that the conformal factor can be seen as coming from a stereographic projection 
of a point on the surface of a 3D sphere onto a 2D plane tangent to it at its south pole. 
Such a stereographic projection is used in 6D in the framework of the IBM, in order to get 
the classical IBM Hamiltonian in terms of the usual collective coordinates, on the 5D plane 
passing through the south pole of the relevant sphere \cite{DSI}.

\subsection{Embedding from five into six dimensions}

The usual Bohr Hamiltonian lives in an euclidean 5D space, the coordinates of which can be labelled as 
\begin{equation}
q_1=\Phi, \quad q_2=\Theta, \quad q_3=\psi, \quad q_4=\beta, \quad q_5=\gamma. 
\end{equation}
These coordinates play the role held by the polar coordinates in the 2D euclidean space, 
or by the spherical coordinates in the 3D euclidean space. 

The line element is \cite{Bohr}
\begin{equation}\label{line}
ds^2 = g_{ij} dq_i dq_j.
\end{equation}
Using the explicit expressions for the matrix elements $g_{ij}$ given in \cite{DDM},
the line element takes the form
\begin{equation}\label{somega4}
ds^2 = d\beta^2 + \beta^2 d\Omega_4^2, 
\end{equation}
where the hyperangular element is 
$$
d\Omega_4^2 =  4 \left[  \sin^2\left(\gamma -{2\pi \over 3} \right)\sin^2\psi +
                  \sin^2\left(\gamma -{4\pi \over 3} \right)\cos^2\psi \right] d\Theta^2  $$ 
                  $$
+ 4 \left[  \sin^2\left(\gamma -{2\pi \over 3} \right)\sin^2\Theta \cos^2\psi 
  +  \sin^2\left(\gamma -{4\pi \over 3} \right)\sin^2\Theta \sin^2\psi +
               \sin^2\gamma \cos^2\Theta \right] d\Phi^2  $$
\begin{equation}               
             + 4 \sin^2\gamma d\psi^2   + 8 \sin^2\gamma \cos\Theta d\Phi d\psi + d\gamma^2
            - 8\sqrt{3} \sin\gamma \cos\gamma \sin\Theta \sin\psi \cos\psi d\Phi d\Theta . \label{omega4}           
\end{equation}             
Eq. (\ref{somega4}) is the 5D analogue of Eq. (\ref{eucl2}) in 2D, and of Eq. (\ref{eucl3}) in 3D,
i.e., the euclidean expression for the line element.  $d\Omega_4^2$, which depends on the four angles 
$\Phi$, $\Theta$, $\psi$, $\gamma$, is the 5D analogue of $d\theta^2$ in 2D (which depends on the single 
angle $\theta$), and of $ d\theta^2 + \sin^2\theta d\phi^2$ in 3D (which depends on the two angles $\theta$, $\phi$). 

The form of Eq. (\ref{somega4}) is in agreement with the fact that the configuration space $R_5$ of the Bohr model
is known \cite{Rowe38,RW} to be the tensor product of a radial line, $R_+$, and a four sphere, $S_4$. 

Letting the nuclear mass $B$ to depend on the deformation $\beta$, it has been proved in \cite{DDM} 
that the matrix elements become 
$ g'_{ij}= g_{ij}/ f^2$,
where $f(\beta)$ is the function expressing this dependence in Eq. (\ref{eq:mBohr}).
As a consequence, the line element of Eq. (\ref{line}) becomes  
\begin{equation}
(ds')^2 = g'_{ij} dq_i dq_j, 
\end{equation}
leading to 
\begin{equation}\label{eucl5}
(ds')^2 = \left( d\beta^2 + \beta^2 d\Omega_4^2\right) f^{-2} , 
\end{equation}
which is the analogue in 5D of Eqs. (\ref{lconf2}) in 2D and (\ref{lconf3}) in 3D. 

In the case of the Davidson potential of Eq. (\ref{Davpot}), in which $f(\beta)$ is given by 
Eq. (\ref{fbeta}), the present Eq. (\ref{eucl5}) becomes 
\begin{equation}\label{dsprime}
(ds')^2 =   \left( d\beta^2 + \beta^2 d\Omega_4^2\right) \left(1+ a \beta^2\right)^{-2} . 
\end{equation}
This result is similar to Eqs. (\ref{lconf2}) and (\ref{lconf3}).
This implies that $\beta$ has resulted from a conformal transformation 
\begin{equation}\label{barb}
\bar \beta = \beta \left( 1+ {\beta^2 \over 4 R^2}\right)^{-1},  
\end{equation}
with $a = 1/(4 R^2)$.

Therefore we see that the DDM parameter $a$ is connected to the radius of curvature $R= 1/(2 \sqrt{a})$ of a hypersphere in 5D,
in agreement with the findings of \cite{Georg}. 
The case of mass independent of the deformation, i.e., $a=0$, corresponds to a hypersphere in 5D of infinite radius, i.e. 
to a hyperplane in 5D. It should be remembered \cite{DSI} at this point that the geometry of the interacting boson model (IBM) \cite{IA}
is that of a 5D space. Therefore, the DDM Bohr results can be compared to the classical limit of the IBM, a task to be undertaken 
in the next section. 

\section{The classical limit of the interacting boson model}

We now turn to the most general IBM Hamiltonian, which reads \cite{IA,GKNPA} 
$$
H = \epsilon_s s^\dagger s + \epsilon_d d^\dagger \cdot \tilde d 
+ u_0 (s^\dagger)^2 s^2 + u_2 s^\dagger d ^\dagger \cdot \tilde d s 
+ v_0 [d^\dagger \cdot d^\dagger s^2 + (s^\dagger)^2 \tilde d \cdot \tilde d] $$
\begin{equation}\label{mostg}
+v_2 [(d^\dagger d^\dagger)^{(2)} \cdot \tilde d s 
 + s^\dagger d^\dagger \cdot (\tilde d \tilde d)^{(2)} ]  
+ \sum_{L=0,2,4} C_L (d^\dagger d^\dagger)^{(L)} \cdot (\tilde d \tilde d)^{L)} , 
\end{equation}
where the symbols preceding the terms involving the boson creation ($s^\dagger$, $d^\dagger$) and annihilation
($s$, $\tilde d = (-1)^\mu d_{-\mu}$) operators are free parameters. 

The corresponding Bohr Hamiltonian, obtained through the appropriate intrinsic state, reads \cite{GKNPA} 
$$
H_B = P_1 {\partial \over \partial \beta} + P_2 {\partial^2 \over \partial \beta^2} +P_{12} {\partial^2 \over 
\partial \beta \partial \gamma} + \bar P_1 {1\over \beta} {\partial \over \partial \gamma} $$
 \begin{equation}\label{mostg2}
+\bar P_2 {1 \over \beta^2 \sin 3\gamma} {\partial \over \partial \gamma} 
\sin 3\gamma {\partial \over \partial \gamma}
+\bar \sum_{k=1,2,3} {\hbar ^2 Q_k^2 \over 2 {\cal J}_k} +V_N, 
\end{equation}
with the explicit expressions of the coefficients appearing here given in terms of the coefficients 
of Eq. (\ref{mostg}) and the total number of bosons $N$ in Ref. \cite{GKNPA}. 

A detailed comparison of Eqs. (\ref{mBohr2}) and (\ref{mostg2}), taking into account the rescaling 
which has been performed in Eq. (\ref{mBohr2}), turns out to be instructive. 

1) Since in Eq. (\ref{mostg2}) the coefficients contain the 9 parameters appearing in Eq. (\ref{mostg}), 
while in relation to Eq. (\ref{mBohr2}) only two parameters enter ($a$, and the rescaling quantity $\hbar^2/B_0$),
it is clear that any attempt to fully map Eq. (\ref{mBohr2}) onto Eq. (\ref{mostg2}) will end up with an overcomplete 
set of equations among the 9 parameters of Eq. (\ref{mostg}).     

2) No terms of the form $\partial /\partial \gamma$ and $\partial^2 \over \partial \beta \partial \gamma$ 
appear in (\ref{mBohr2}). Since the relevant coefficients in (\ref{mostg2}), 
\begin{equation}\label{barP1}
\bar P_1 = \sqrt{2\over 7}  v_2 \left( 1+ {3N \beta^2 \over 1+\beta^2}   \right)\sin 3\gamma, 
\end{equation}
\begin{equation}\label{P12}
P_{12}=  \sqrt{2\over 7} v_2 (2-\beta^2) \sin 3\gamma, 
\end{equation}
 are proportional to $\sin 3\gamma$ \cite{GKNPA},
it turns out that the DDM Hamiltonian (\ref{mBohr2}) is missing some triaxiality effects contained in the classical limit 
of the most general IBM Hamiltonian (\ref{mostg2}). Since these coefficients are also proportional to $v_2$, we are 
going to ignore terms proportional to $v_2$ in what follows, formally considering $v_2=0$. 

3) The coefficients of the term $(\sin 3\gamma)^{-1} (\partial / \partial \gamma) 
\sin 3\gamma (\partial/ \partial \gamma)$ \cite{GKNPA} lead to 
\begin{equation} 
\bar P_2= v_0 + \left( {1\over 5} C_0 - {2\over 7} C_2 + {3\over 35} C_4 \right) \beta^2 + \sqrt{2\over 7} v_2 \beta \cos 3\gamma  = -{\hbar^2\over 2 B_0} (1+ a \beta^2)^2, 
\end{equation}
from which the following remarks follow. 

a) As above, there is a term $\beta \cos 3\gamma$ in (\ref{mostg2}), arising from the $v_2$ term in (\ref{mostg}). 
As discussed in 2), no such term exists in  (\ref{mBohr2}). 

b) A $\beta^4$ term arises in (\ref{mBohr2}), but not in (\ref{mostg2}). In other words, the DDM Bohr Hamiltonian 
contains some higher order terms not present in the classical limit of the IBM Hamiltonian. On the other hand, these 
higher order terms are preceded by $a^2$, which is a small quantity since $a<<1$ \cite{DDM}. 

c) The coefficients of $\beta^2$ lead to
\begin{equation}\label{aIBM}
\tilde a= -{1\over 5} C_0 +{2\over 7} C_2 - {3\over 35} C_4,
\end{equation}
where 
\begin{equation}
\tilde a = {\hbar^2 \over B_0} a.
\end{equation}
thus connecting the DDM parameter $a$ to certain IBM parameters. 

4) The coefficients of the terms containing the components of the angular momentum, $Q_k$, 
\begin{equation}
{\hbar^2 \over 2 {\cal J}_k} = -{1\over 7} (C_2-C_4) -
 { v_0 + \left( {1\over 5} C_0 - {2\over 7} C_2 + {3\over 35} C_4 \right) \beta^2 
+ \sqrt{2\over 7} v_2 \beta \cos\left( \gamma -{2\pi k\over 3}  \right)   \over
 4\beta^2 \sin^2 \left(  \gamma -{2\pi k\over 3} \right)  },
\end{equation}
lead to the same results as in 3).  

The results up to now can be summarized as follows.

1) The DDM Bohr Hamiltonian contains some higher order terms not present in the 
IBM Hamiltonian. 

2) On the contrary, the classical limit of the IBM Hamiltonian 
contains $\gamma$-dependent terms absent from the DDM Bohr Hamiltonian.  

3) It is interesting to see what are the implications of Eq. (\ref{aIBM}) 
in the three limiting symmetries of IBM.

Before proceeding to the study of the three limiting symmetries of the IBM, a comment on terms 
allowed in the general form of the Bohr collective model is in place. In general, terms of the form 
$\beta^i (\cos 3\gamma)^j $ can occur \cite{Maruhn}. The term $i=0$ is always excluded, since
for $j\neq 0$ it  implies an indefinite value for the potential at $\beta=0$, while for $j=0$ it is just 
a constant \cite{Maruhn,Hess}. The well known 24 transformations which have to leave the wave function invariant 
\cite{Bohr,Corrigan} imply that even values of $i$ should be accompanied by even values of $j$, or $j=0$, 
while odd values of $i$ should be accompanied by odd values of $j$. Usually, $i=1$ is not included in the  
potentials used in the collective model \cite{Hess,Hess2,Gneuss}, since it does not have a smooth behaviour at 
$\beta=0$, but it could be included \cite{Hess}, since it does not violate any symmetry constraints. 
Therefore the appearance of the term $\beta \cos 3\gamma$ in the above discussion is not problematic.  

\subsection{The U(5) limit}

In the U(5) (vibrational) limit a simple IBM Hamiltonian 
can be written as \cite{IA,GKNPA}
\begin{equation}
H_{\rm U(5)} = \epsilon_d d^\dagger \cdot \tilde d + \kappa_5 d^\dagger \cdot d^\dagger  \tilde d \cdot \tilde d.
\end{equation}
In other words, the only non-vanishing coefficients in (\ref{mostg}) are $\epsilon_d$ and $C_0=5 \kappa_5$. 
Then Eq. (\ref{aIBM}) yields 
\begin{equation}
\tilde a = - \kappa_5. 
\end{equation}
Thus in the vibrational limit the DDM parameter $a$ turns out to be related 
to the strength of the pairing interaction among the $d$-bosons. It should be pointed out that 
$\kappa_5$ obtains negative values (see Eqs. (4.11a) and (4.12) of Ref. \cite{GKNPA}), 
thus guaranteeing that $a>0$, as it should be \cite{DDM}.  

A more general IBM Hamiltonian in the U(5) limit reads \cite{IA,AIU5}
\begin{equation}
H_{\rm U(5)} = \epsilon_d d^\dagger \cdot \tilde d + \sum_{L=0,2,4} C_L (d^\dagger d^\dagger)^{(L)} \cdot  (\tilde d \tilde d)^{(L)},
\end{equation}
in which the non-vanishing coefficients in (\ref{mostg}) are $\epsilon_d$, $C_0$, $C_2$, $C_4$. 
In this case, $\tilde a$ is given by Eq. (\ref{aIBM}). 

\subsection{The SU(3) limit} 

We assume now a quadrupole--quadrupole Hamiltonian of the form \cite{IA,AISU3}
\begin{equation}
H_{\rm{SU(3)}} = -\kappa Q\cdot Q - \kappa' L \cdot L  
\end{equation}
where $Q$ is the quadrupole operator 
\begin{equation}
Q= s^\dagger \tilde d_\mu + d^\dagger_\mu s +\chi (d^\dagger \tilde d)^{(2)}_\mu,
\end{equation}
and $L$ is the angular momentum operator
\begin{equation}\label{LL}
L=\sqrt{10} (d^\dagger \tilde d)^{(1)}. 
\end{equation}
 
This choice of the Hamiltonian corresponds to specific values of the parameters
of Eq. (\ref{mostg}) in terms of $\kappa$ and $\kappa'$, including \cite{AISU3,GKNPA}
\begin{equation}
C_0= -{7\over 4} \kappa + 6 \kappa', \quad C_2 = {3\over 8} \kappa + 3 \kappa', \quad  C_4= -{1\over 2} \kappa -4 \kappa',
\end{equation}
in which the differences in the definitions of the parameters in Table I of Ref. \cite{AISU3} and in Eq. (5.19) of Ref. \cite{GKNPA}
have been taken into account, as shown in the Appendix A1.1~.  
These parameters, used in Eq. (\ref{aIBM}), lead to 
\begin{equation}
\tilde a = {1\over 2} \kappa. 
\end{equation}
Thus in the SU(3) limit the DDM parameter $a$ turns out to be related to the strength 
of the quadrupole-quadrupole interaction, while the coefficient of the angular momentum term 
does not enter. It is known \cite{AISU3} that $\kappa$ takes positive values, thus 
guaranteeing $a>0$, as it should be \cite{DDM}. 

\subsection{The O(6) limit} 

In the O(6) ($\gamma$-unstable) limit the most general Hamiltonian takes the form \cite{IA,AIO6}
\begin{equation}
H_{\rm O(6)}= A P_6+ B C_5 + C L\cdot L, 
\end{equation}
where $L$ is given by Eq. (\ref{LL}), and 
\begin{equation}
P_6 = P^\dagger \cdot \tilde P, \qquad \tilde P = {1\over 2} (\tilde d \cdot \tilde d)-{1\over 2} (\tilde s \cdot \tilde s), 
\qquad C_5 = {1\over 3} \sum_{1,3} (d^\dagger \tilde d)^{(k)} \cdot (d^\dagger \tilde d)^{(k)}. 
\end{equation}
This choice of the Hamiltonian corresponds to specific values of the parameters
of Eq. (\ref{mostg}) in terms of $A$, $B$, $C$,  including \cite{AIO6}
\begin{equation}
C_0= {5\over 4} A -{2\over 3} B -6 C, \quad C_2 = {1\over 6} B-3C, \quad  C_4= {1\over 6}B +4C,
\end{equation}
in which the differences in the definitions of the parameters in Table I of Ref. \cite{AIO6} and in Ref. \cite{GKNPA}
have been taken into account, as shown in the Appendix A1.2~.   
These parameters, used in Eq. (\ref{aIBM}), lead to 
\begin{equation}
\tilde a = -{1\over 4} A +{1\over 6} B. 
\end{equation}
Thus in the O(6) limit the DDM parameter $a$ turns out to be related to the difference of the strengths
of a pairing interaction involving both $s$ and $d$ bosons and a pairing interaction involving $d$ bosons only,   
while the coefficient of the angular momentum term again does not enter. 

\subsection{Discussion}

We have found that in all the U(5), SU(3), and O(6) limits the DDM parameter $a$ is connected 
to relevant IBM parameters.

It should be remarked that the very presence of non-vanishing $d$-boson pairing interaction
in vibrational nuclei, or of non-vanishing quadrupole-quadrupole interaction in rotational nuclei,
or of non-vanishing difference between the pairing interactions among $s$, $d$ bosons and $d$ bosons alone 
in $\gamma$-unstable nuclei, 
requires a non-vanishing value of the DDM parameter $a$, implying a finite curvature of the 5D Bohr space.
In other words, the classical limit of the IBM Hamiltonian corresponds to a curved 5D space \cite{GKPRL}, 
while in the original Bohr Hamiltonian the 5D space is flat (corresponding to $a=0$, i.e., to infinite 
radius of curvature). 
  
The connection between curvature and interaction appears in several branches of physics.

1) In the general theory of relativity \cite{Einstein}, the Einstein field equation (actually a set of 10 equations) 
connects the local spacetime curvature
(expressed by the Einstein tensor, constructed from the Riemannian curvature tensor and the metric)
to the local energy and momentum within that spacetime (expressed by the stress-energy tensor, defined by the matter content of the spacetime), thus describing quantitatively the principle that ``spacetime tells matter how to move, and matter 
tells spacetime how to curve'' \cite{Wheeler}.

2) In thermodynamics, the Ruppeiner geometry has been developed \cite{Rupp1,Rupp2}, in which thermodynamical 
systems are represented in terms of Riemannian geometry, the relevant metric being flat for noninteracting particles, 
while curvature develops in the presence of interactions \cite{Quevedo}.      
  
Connections between curvature and interaction have also been considered in the realm of nuclear structure.

1) In the search for a collective path in the many-particle Hilbert space, the manifold of Slater determinants has been 
considered as a Riemannian manifold \cite{Giraud}, with the curvature of a collective path, expressed through 
an external curvature tensor, found to be related to its collectivity. In particular, geodesics (lines of zero internal curvature) on generic SU(3) orbits are found to be highly collective paths \cite{Giraud}.     
  
2) By interpreting geometrically the collective masses, the metric tensor can be defined, which fully determines 
the geometric properties of the collective Riemannian  space \cite{Herrmann}. Assuming non-vanishing curvature in collective space, the finite range liquid drop model with curvature, and the finite range droplet model with curvature, have been 
constructed, giving improved ground state properties for light nuclei and trans-fermium elements \cite{Herrmann}.    

In view of the above comments, the present results can be interpreted in the following way. The curvature 
of the space is determined by the leading interaction present in each symmetry limit of the IBM. Since the curvature is directly 
related to the free parameter $a$ appearing in the dependence of the mass on the deformation, it turns out that 
the parameter $a$ is also determined by the leading interaction present in each symmetry limit. 
  
\section{Conclusions}

The physical meaning of the free parameter $a$ appearing in the dependence of the mass on the deformation 
in the Bohr Hamiltonian with a Davidson potential has been considered. The main results are summarized here.

1) By embedding the 5D DDM Bohr space into a 6D space, the parameter $a$ has been connected to the curvature 
of the 5D space, the original Bohr Hamiltonian corresponding to a flat 5D space. 

2) By comparing the deformation-dependent mass (DDM) Bohr Hamiltonian to the classical limit 
of the most general IBM Hamiltonian, the parameter $a$ has been connected to certain IBM parameters.
In particular the parameter $a$ has been found a) in the U(5) limit to be proportional to the strength 
of the $d$-boson pairing interaction, b) in the SU(3) limit to be proportional to the strength 
of the quadrupole-quadrupole interaction, c) in the O(6) limit to be proportional to a difference of the strengths 
of the pairing interaction among $s$ and $d$ bosons and the pairing interaction among $d$-bosons alone.   

3) The presence of certain basic interactions in the IBM results in a curved 5D space corresponding to its classical limit, 
while the 5D space of the original Bohr Hamiltonian is a flat one.
Curvature needs to be added to the 5D space of the Bohr Hamiltonian, by allowing the nuclear mass to depend on the deformation, 
in order to establish agreement with the classical limit of IBM.   
In other words, the IBM in its classical limit, as already remarked in Ref. \cite{GKNPA},
has built-in the dependence of the nuclear mass on the deformation,
which has been introduced in the DDM Bohr Hamiltonian in order to fix the behaviour of the moments of inertia 
as functions of the deformation. 

The influence of the parameter $a$, i.e. of the curvature of the 5D space, 
on the properties of critical point symmetries \cite{IacE5,IacX5} and shape phase transitions 
\cite{McCutchan,RMP82} in atomic nuclei is an interesting problem to be pursued.
The embedding of the Bohr space in 6D has already been used in revealing the O(6) symmetry 
and its contraction to the E(5) symmetry at infinity \cite{Georg}.   

The DDM approach to the Bohr Hamiltonian with a Kratzer potential \cite{Kratzer}  
has been recently carried out \cite{DDMK}, providing a different factor $f(\beta)=1 + \alpha \beta$
(with $a <<1$). The extension of the present embedding approach to the Kratzer case is an interesting task,
which might require the use of a different coherent state.   

\section*{Acknowledgements}

The authors are thankful to P. E. Georgoudis for useful discussions.  
Financial support from the Bulgarian National Science Fund under contract No. DFNI-E02/6 
and by the Scientific Research Projects Coordination Unit of Istanbul University under Project No 50822
is gratefully acknowledged.

\section*{Appendix: Coefficients of different Hamiltonians}

In the most general IBM Hamiltonian given in Eq. (\ref{mostg}) as written in Ref. \cite{GKNPA}, 
there are some differences in the coefficients in comparison to the most general IBM Hamiltonian 
reported in Ref. \cite{IA}
$$
H = \epsilon_s s^\dagger s + \epsilon_d d^\dagger \cdot \tilde d 
+ {1\over 2} u_0 (s^\dagger)^2 s^2 + u_2 [(s^\dagger d ^\dagger)^{(2)} \times (\tilde d s)^{(2)}]^{(0)}  $$
$$ + {1\over 2} v_0 [(d^\dagger d^\dagger)^{(0)} \times  (ss)^{(0)} 
+ (s^\dagger s^\dagger)^{(0)} \times (d \tilde d)^{(0)}]^{(0)} $$
$$ +v_2 [(d^\dagger d^\dagger)^{(2)} \times (\tilde d s)^{(2)} 
+ (s^\dagger d^\dagger)^{(2)} \times (\tilde d \tilde d)^{(2)} ]^{(0)} $$
\begin{equation}\label{mostgIA}
+ \sum_{L=0,2,4} {1\over 2} \sqrt{2L+1} C_L [(d^\dagger d^\dagger)^{(L)} \times (\tilde d \tilde d)^{L}]^{(0)} . 
\end{equation}
Using the identity
\begin{equation}
U^{(k)} \cdot V^{(k)}=(-1)^k \sqrt{2k+1} (U^{(k)} V^{(k)})^{(0)},
\end{equation}
one easily finds the following relations between the Iachello--Arima parameters of Eq. (\ref{mostgIA}),
labelled by IA, 
and the Ginocchio--Kirson parameters of Eq. (\ref{mostg}), labelled by GK
$$
\epsilon_{s,GK} = \epsilon_{s,IA}, \quad \epsilon_{d,GK} = \epsilon_{d,IA}, \quad
\sqrt{5} v_{0,GK}= {1\over 2} v_{0,IA}, $$
$$
 \sqrt{5} v_{2,GK}= {1\over \sqrt{2}} v_{2,IA},\quad
u_{0,GK}= {1\over 2} u_{0,IA}, \quad
 \sqrt{5} u_{2,GK}= u_{2,IA}, $$ 
 \begin{equation}\label{GKIA}
C_{0,GK}={1\over 2} C_{0,IA}, \quad C_{2,GK}={1\over 2} C_{2,IA},\quad 
 C_{4,GK}={1\over 2} C_{4,IA}.  
\end{equation}

\subsection*{A1.1 SU(3)}

From Table I of Ref. \cite{AISU3} one reads 
$$ 
\epsilon_{s,IA}= -5\kappa, \epsilon_{d,IA}= -{11\over 4} \kappa -6 \kappa',  \quad
 v_{0,IA}=-2\sqrt{5}\kappa, \quad v_{2,IA}=\sqrt{70}\kappa,  \quad
u_{0,IA}=0, $$
\begin{equation}
 u_{2,IA} = -2\sqrt{5}\kappa, \quad
C_{0,IA}= -{7\over 2} \kappa +12 \kappa', \quad 
C_{2,IA}= {3\over 4}\kappa +6 \kappa', \quad
C_{4,IA}= -\kappa -8\kappa'. 
\end{equation}
Taking into account the relations of Eq. (\ref{GKIA}) one finds
$$
\epsilon_{s,GK}= -5\kappa, \epsilon_{d,GK}= -{11\over 4} \kappa -6 \kappa',  \quad
 v_{0,GK}=-\kappa, \quad v_{2,GK}=\sqrt{7}\kappa,  \quad
u_{0,GK}=0, $$
\begin{equation}
 u_{2,GK} = -2\kappa,  \quad
 C_{0,GK}= -{7\over 4} \kappa +6 \kappa', \quad C_{2,GK}= {3\over 8}\kappa +3 \kappa', \quad 
 C_{4,GK}= -{1\over 2}\kappa -4\kappa', 
\end{equation}
in agreement to Eq. (5.19) of Ref. \cite{GKNPA} as far as the $\kappa$ terms are concerned. 

\subsection*{A1.2 O(6)}

From Table I of Ref. \cite{AIO6} one reads 
$$
\epsilon_{s,IA}= 0, \epsilon_{d,IA}= {2\over 3} B +6 C, \quad
\quad v_{0,IA}=-{\sqrt{5}\over 2}A, \quad v_{2,IA}=0, \quad
u_{0,IA}={1\over 2}A, $$ 
\begin{equation}
 u_{2,IA} = 0,  \quad
 C_{0,IA}= {5\over 2} A -{4\over 3} B-12 C, \quad C_{2,IA}= {1\over 3} B -6 C, \quad
 C_{4,IA}= {1\over 3}B +8 C. 
\end{equation}
Taking into account the relations of Eq. (\ref{GKIA}) one finds
$$
\epsilon_{s,GK}= 0, \epsilon_{d,GK}={2\over 3} B +6 C,  \quad
 v_{0,GK}=-{1\over 4}A, \quad v_{2,GK}=0,  \quad
u_{0,GK}={1\over 4}A, $$ 
\begin{equation}
u_{2,GK} = 0, \quad
 C_{0,GK}={5\over 4} A -{2\over 3} B-6 C, \quad C_{2,GK}= {1\over 6} B -3 C, \quad
 C_{4,GK}= {1\over 6}B +4 C.
\end{equation}

\end{document}